\documentclass[prb,english,twocolumn,amsmath,amssymb,superscriptaddress]{revtex4-2}
\usepackage{amsmath}
\usepackage{graphicx}
\usepackage{amssymb}
\usepackage{dsfont}
\usepackage{color}
\usepackage[bookmarks=true,colorlinks,linkcolor=blue,urlcolor=blue,citecolor=blue]{hyperref}

\usepackage{hyperref}
\usepackage{tikz}
\usetikzlibrary{calc}

\newcommand{\be}{\begin{equation}}
\newcommand{\ee}{\end{equation}}
\newcommand{\bea}{\begin{eqnarray}}
\newcommand{\eea}{\end{eqnarray}}
\newcommand{\ket}{\rangle}
\newcommand{\bra}{\langle}

\definecolor{darkorange}{rgb}{1, 0.55, 0.0}

\begin{document}

\title{Coherence Properties of the Repulsive Anyon-Hubbard Dimer}

\author{Martin Bonkhoff}
\affiliation{Physics Department and Research Center OPTIMAS,
	University of Kaiserslautern-Landau, 67663 Kaiserslautern, Germany}
\author{Simon B. Jäger}
\affiliation{Physics Department and Research Center OPTIMAS,
	University of Kaiserslautern-Landau, 67663 Kaiserslautern, Germany}
\author{Imke Schneider}
\affiliation{Physics Department and Research Center OPTIMAS,
	University of Kaiserslautern-Landau, 67663 Kaiserslautern, Germany}
\author{Axel Pelster}
\affiliation{Physics Department and Research Center OPTIMAS,
	University of Kaiserslautern-Landau, 67663 Kaiserslautern, Germany}
\author{Sebastian Eggert}
\affiliation{Physics Department and Research Center OPTIMAS,
	University of Kaiserslautern-Landau, 67663 Kaiserslautern, Germany}

\begin{abstract}
 
One-dimensional anyonic models of the Hubbard type show intriguing ground-state properties, effectively transmuting between Bose-Einstein and Fermi-Dirac statistics. 
The simplest model that one can investigate is an anyonic version of the bosonic Josephson junction, the repulsive anyon-Hubbard dimer. 
In the following we find an exact duality relation to the Bethe-solvable Bose-Hubbard dimer, which is well known from quantum optics and information theory and has interesting connections to spin squeezing and entangled coherent states. Conversely, we show that the anyonic Hubbard dimer has non-trivial coherence properties that emerge from the anyonic statistics. In particular, we find that coherences can be suppressed and amplified and show that these features are remarkably robust against additional repulsive onsite interactions highlighting the distinct nature of anyons.

\end{abstract}

\date\today

\maketitle

\section{Introduction}
Recent progress in the experimental realization of one-dimensional anyons 
via density-dependent gauge-phases \cite{Froelian2022,Chisholm2022,Kwan:2023} 
warrants a deeper theoretical understanding of these non-standard statistical interactions. 
In one spatial dimension the concept of anyons does not arise from the same topological origin as in the two-dimensional world \cite{Pauli1925,Leinaas,Wilczek,Harshman2022,Lundholm:2023}. Nevertheless, one-dimensional anyons can be defined via a fractional Jordan-Wigner transformation to simulate the exchange phases of two-dimensional abelian anyons structurally. Thereby the string operator is attached to a bosonic or fermionic parent particle, whose nature dictates the local exclusion behavior of the anyons. 
For bosonic parent particles, the resulting anyon Hubbard models have been shown to possess intriguing superfluid properties \cite{Bonkhoff2021}, alter the parameter regions of phase transitions \cite{Keilmann2011}, or even stabilize new phases of matter under statistical transmutation \cite{Lange2017a}. 

In this paper we consider the simplest version of such lattice anyons, namely the anyonic  
Hubbard dimer defined on only two lattice sites. 
The dynamics of this model has been explored in the context of Josephson Junctions using a mean-field coherent state approach~\cite{Brollo2022}. We now show that the anyonic Hubbard dimer can be mapped by a duality relation to the exactly solvable Bose-Hubbard dimer~\cite{Enolskii1993,Jon2006}, which in turn allows a systematic analysis of the coherence ~\cite{Glauber1963} and the corresponding experimentally measurable correlation functions. In previous works, 
 assisted Raman tunneling and shaking were proposed to induce a density-dependent 
complex phase in the hopping elements. This method may allow the experimental
 realization of anyonic 
physics~\cite{Keilmann_2011, Greschner, Tang, Straeter2016}, 
where a fast time-periodic modulation of the 
interaction~\cite{Ramos_2008, Pollack_2010} will lead to an 
effective hopping matrix element 
depending on the density difference~\cite{Rapp_2012, Wang_2014, Greschner_2014, Eckardt_2017, Arimondo_2012, Meinert_2016, Wang_2020}.
Our proposed unitary transformation realizes density-dependent tunneling without fast time-periodic driving. Instead, it is based on non-linear elements which are well known
in the field of quantum optics,  embodying certain spin squeezing transformations \cite{Kitagawa1993},  optical Kerr non-linearities  \cite{Kitagawa1986}, or non-linear interferometers \cite{Sanders1999,Rice2000}. Our findings do not only provide a new way to realize a particular version of the anyon-Hubbard model  \cite{Keilmann2011,Longhi2011,Longhi2012,Greschner,Straeter2016,Yuan2017,Schweizer2019,Goerg2019}, but conversely describe its unconventional coherence properties. 
In particular, we find for large enough $N$ that the statistical interactions in the dimer model correspond to filters or amplifiers of multi-particle excitations. The fate of these excitations depends on the particular value of the statistical parameter and the exact number of excited particles.  To this end we proceed as follows. After introducing the underlying anyon-Hubbard model in Sec.~\ref{sec:model} the duality relation is found in Sec.~\ref{sec3}.
A central role in this work plays thereby the calculation of the $m$-th order, off-diagonal  $g$-function \cite{Glauber1963,Albiez2005,Gati2006,Gati2007,Ferrini2008,Mazzarella2011} introduced in Sec.~\ref{sec4}.
First, we employ a mean-field approximation for large $N$ using a Heisenberg-Weyl coherent state to calculate the coherence function in Sec.~\ref{sec:meanfield}. Subsequently, the case of zero onsite repulsion, $U=0$, is treated in Sec.~\ref{sec:squeezing}, where the $m$-th order $g$-function is obtained exactly with the help of non-linear $su(2)$ coherent states~\cite{Wang2000}. With this we determine the resonance conditions of minimal and maximal coherence which highlight the filtering and amplification of multi-particle excitations. The role of $U>0$ is analyzed in Sec.~\ref{sec:squeezing2} where we show that effects emerging from statistical interactions are remarkably robust against $U$. Finally, we conclude our results in Sec.~\ref{sec:Conc}.

\section{The Model\label{sec:model}}

One-dimensional anyons of the Hubbard type  are defined by a set of deformed commutation relations for their creation and annihilation operators $\hat{a}^{\dagger}_j,\hat{a}_i$ at lattice sites $j,i$ \cite{Keilmann2011}, i.e.
\begin{align}
	&\hat{a}_i\hat{a}^{\dagger}_j-e^{i\theta\mathrm{sgn}(i-j)}\hat{a}^{\dagger}_j	\hat{a}_i=\delta_{i,j}\nonumber\\
 &\hat{a}_i\hat{a}_j-e^{-i\theta\mathrm{sgn}(i-j)}\hat{a}_j	\hat{a}_i=0. \label{deformedcommutator1}
\end{align}
The deformed commutators in Eq.~(\ref{deformedcommutator1}) depend on the statistical parameter $\theta\in\left[0,\pi\right]$ and on the spatial ordering of the lattice via the sign function $\mathrm{sgn}(x)=x/\vert x\vert$ with convention $\mathrm{sgn}(0)=0$. Equation~\eqref{deformedcommutator1} allows for  considering the Hubbard anyons in bosonic formulation with a fractional Jordan-Wigner transformation \cite{Keilmann2011},
\begin{align}
\hat{a}_{j}=\hat{b}_{j}e^{i\theta\sum_{l<j}\hat{n}_l}\label{jw},\quad \hat{n}_j=\hat{a}^{\dagger}_j\hat{a}_j=\hat{b}^{\dagger}_j\hat{b}_j,
\end{align}  
mediating between the anyons and the bosonic parent particle, i.e.~$\left[\hat{b}_{i},\hat{b}^{\dagger}_{j}\right]=\delta_{i,j}$. The Hamiltonian of the anyonic dimer model in bosonic representation is given by \cite{Brollo2022}, i.e.
\begin{align}
	\hat{H}=-2J\left(\hat{b}^{\dagger}_1\hat{b}_2e^{i\theta\hat{n}_1}+\mathrm{h.c.}\right)+\frac{U}{2}\sum_{i=1}^{2}\hat{n}_i(\hat{n}_i-1)\label{Hamiltonian}.
\end{align} 
This Hamiltonian describes density-dependent tunneling between the two sites with rate $J$ and onsite interactions that we assume to be repulsive $U>0$ in this paper. In the upcoming Sec. \ref{sec3}, we show that $\hat{H}$ [Eq.~\eqref{Hamiltonian}] for any $\theta$ is unitary equivalent to $\hat{H}$ for $\theta=0$ which is the Bose-Hubbard dimer Hamiltonian.

\section{Duality Relation  \label{sec3}}

The $N$-particle Hamiltonian in Eq.~(\ref{Hamiltonian}) can be conveniently represented in terms of its $su(2)$ current algebra \cite{Enolskii1993,Bogoliubov2016,Jon2006,Lipkin1965},
\begin{align}
	\label{Lipkinmeshkovglick}
	\hat{H}=-2J\left(\hat{s}^+\vert_\theta+\hat{s}^-\vert_\theta\right)+U\left(\hat{s}^z\vert_\theta\right)^2+C_N.
\end{align}
The last term $C_N=U(N/2-1)N/2$ only depends on the constant particle number and this is why it is unimportant for our further investigations. The currents in Eq.~\eqref{Lipkinmeshkovglick} are represented by bi-linear combinations of anyonic operators $\hat{a}_j$ defined as
\begin{align}
	&\hat{s}^+\vert_\theta=\hat{a}^{\dagger}_1\hat{a}_2=\hat{b}^{\dagger}_1\hat{b}_2e^{i\theta\hat{n}_1},\label{anyonicladdersu2}
	\\
	&\hat{s}^{z}\vert_\theta=\frac{1}{2}\big(
	\hat{n}_1-\hat{n}_2\big)=\hat{s}^z\vert_{\theta=0}\label{anyoniclongitudinalsu2}.
\end{align} 
Those operators obey the spin commutation rules $[\hat{s}^+\vert_\theta,\hat{s}^-\vert_\theta]=2\hat{s}^z\vert_\theta$, $\hat{s}^-\vert_\theta=(\hat{s}^+\vert_\theta)^\dag$ and $[\hat{s}^z\vert_\theta,\hat{s}^+\vert_\theta]=\hat{s}^+\vert_\theta$ such that they can be seen as $su(2)$ spins originating from bosons with an attached gauge phase \cite{Jordan1935}. In this language the number of particles corresponds to the conserved total spin magnitude $\hat{N}/2=(\hat{n}_1+\hat{n}_2)/2$. We demonstrate now that the anyonic currents ${\hat{s}^{\pm,z}\vert_\theta=\hat{U}\hat{s}^{\pm,z}\vert_{\theta=0}\hat{U}^\dag}$ are related by a similarity transform $\hat{U}$ to the linear, bosonic ones $\hat{s}^{\pm,z}\vert_{\theta=0}$ for all $\theta$. This unitary transformation between the bosonic and anyonic operators must be non-linear
as it maps quadratic to non-quadratic operators. It can be expressed as
\begin{align}
	\hat{U}(\theta)=e^{i\frac{\theta}{2}\hat{n}_1(\hat{n}_1-1)}
\label{duality}
\end{align}
and transforms the boson creation operators
\begin{align}
	&\hat{U}\hat{b}^{\dagger}_1\hat{U}^{\dagger}
=\hat{b}^{\dagger}_1e^{i\theta\hat{n}_1},\label{inversedual1}
\end{align}
where we have used $f(\hat{n}_1)\hat{b}_1^\dag=\hat{b}_1^\dag f(\hat{n}_1+1)$ for an arbitrary function $f$ of the operator $\hat{n}_1$. 
Note, that we could have expressed the unitary operator also in terms of spin operators by using $\hat{n}_1=\hat{s}^z+\hat{N}/2$. Non-linear transformations involving the conserved total particle number $\hat{N}^2$ do not change the model but do not commute with the bosonic operators.  Therefore, non-linear transformations involving the $\propto\hat{s}^z\hat{s}^z$ spin-interaction may also be used to realize the anyonic model. Such transformations are non-local and are related to an effect called one-axis twisting~\cite{Kitagawa1993,Pezze:2018} that can be used for spin squeezing. 

The existence of a unitary transformation $\hat{U}$ between the Bose dimer and the anyon dimer implies that the corresponding Hamiltonians are iso-spectral. 
As a direct consequence, since $\hat{U}$ in Eq.~\eqref{duality} commutes with the local densities $\hat{n}_1$ and $\hat{n}_2$, we find the equivalence of observables that are purely density-dependent and the invariance of the ground-state entanglement entropy with respect to $\theta$ as shown in Ref.~\cite{Brollo2022}. Moreover, the ground-state 
\begin{align}
\vert{\gamma}( \theta)\ket=\hat{U}(\theta)\vert{\gamma}(0)\ket\label{Howtogetthegroundstate}
\end{align}
 of the anyonic dimer can be obtained by applying $\hat{U}$ to the ground-state of the Bose-Hubbard dimer $\vert{\gamma}(0)\ket$. 
 
 From a fundamental point of view, the unitary $\hat{U}$ is a bond-algebraic duality according to Ref.~\cite{Cobanera2011a}, which does not only map two Hamiltonians into each other but also their bonds, embodied by the two different $su(2)$ algebras $\hat{s}^{\pm,z}\vert_{\theta}$ and $\hat{s}^{\pm,z}\vert_{\theta=0}$. Non-linear $su(2)$ spins, gauge related to those in Eqs.~\eqref{anyonicladdersu2} and \eqref{anyoniclongitudinalsu2} were investigated in the context of mutually unbiased bases and $su(2)$ phase states \cite{Atakishiyev2010a}, as well as non-linear coherent states \cite{Wang2000}. The anyonic transversal currents $\hat{s}^{x,y}\vert_{\theta}$ are implicitly periodic in $\theta$, manifested by a generalized Condon-Shortley phase \cite{Atakishiyev2010a}, in accordance with the topological character of the anyonic exchange~\cite{Bonkhoff2021}. Moreover, since the Bose-Hubbard dimer is exactly solvable~\cite{Enolskii1993,Jon2006}, this also implies that the anyon-Hubbard dimer is solvable via the algebraic Bethe ansatz and the duality relation~\eqref{duality} (see App.~\ref{bethesolution}). 
 
From an experimental point of view, the existence of a 
unitary transformation in Eq.~\eqref{duality} paves the path to 
study anyonic models by 
designing 
local or non-local, non-linear operations 
on different bosonic systems including, e.g., photons and macroscopic spins. For the generation of entangled coherent states of photons, such transformations have been described in Refs.~\cite{Sanders1999,Rice2000} via non-linear
Mach-Zehnder interferometry. The general idea is that photons interact over a finite time $\tau$ with a Kerr medium leading to the described $\hat{U}(\theta)$ with $\theta\propto \tau$~\cite{Kitagawa1986}. Alternatively, one can also realize such non-linear gates using a large ensemble of two-level atoms that form a macroscopic spin. In such setups $\hat{U}$ is known as a spin squeezing transformation~\cite{Kitagawa1993} and can be realized by coupling the atoms collectively to an optical cavity~\cite{Carrasco:2022}. To be specific, Eq.~\eqref{Howtogetthegroundstate} means if one can realize the ground-state of the Bose-Hubbard dimer in a lab and apply a non-linear quantum gate in the form of $\hat{U}$ one has automatically realized the ground-state of an anyonic dimer. This finding suggests that $\vert{\gamma}(\theta)\ket$ for specific values of $\theta$ has likely already been realized in experimental labs~\cite{Pedrozo:2020,Greve:2022} without identifying the physical state with the ground-state of the anyonic Hubbard dimer.

The duality relation of both models leads to an equivalent physical interpretation of static or dynamical dual observables, which implies in our context the invariance of Josephson physics under statistical transmutation  \cite{Albiez2005,Gati2006,Gati2007,Brollo2022}.  Nevertheless, the experimentally measurable coherence properties of Eq.~\eqref{Hamiltonian} are known to be related to the bosonic transversal currents $\hat{b}_1^\dag\hat{b}_2=\hat{s}^{\pm}\vert_{\theta=0}$, not to the anyonic ones in Eq.~\eqref{anyonicladdersu2} \cite{Keilmann2011,Greschner,Tang}. Analogous to the calculation of transversal spin-$1/2$ correlation functions via fermionization \cite{Lieb1961}, these observables are not protected by duality and thus become non-trivially dependent on the statistical parameter $\theta$. As a consequence one can ask the question how $\theta$ affects the currents which is at the center of investigation in the following section.    

\section{Coherence\label{sec4}}  
To quantitatively analyze the properties of the currents $\hat{b}_1^\dag\hat{b}_2=\hat{s}^{\pm}\vert_{\theta=0}$ we introduce the $m$-th order coherence function~\cite{Glauber1963}
\begin{align}
	\label{gfunction}
	g^{(m)}(\theta)=\frac{\langle(\hat{b}_1^\dag\hat{b}_2)^m\rangle}{\sqrt{\langle\hat{n}_1\rangle^m\langle\hat{n}_2\rangle^m}}=\frac{\langle \left(\hat{s}^{+}\vert_{\theta=0}\right)^m\rangle}{(N/2)^m}.
\end{align}  
Here the expectation value $\bra\hat{O}\ket=\bra\gamma(\theta)\vert\hat{O}\vert\gamma(\theta)\ket$  of an operator $\hat{O}$ is taken over the ground-state $\vert\gamma(\theta)\ket$ of the anyonic dimer Hamiltonian~\eqref{Lipkinmeshkovglick}. Since the Bose-Hubbard dimer Hamiltonian is invariant under the exchange of site 1 and site 2 and $\hat{U}(\theta)$ does not modify the local densities, we can use $\langle\hat{n}_1\rangle=N/2=\langle\hat{n}_2\rangle$ which shows the second equal sign in Eq.~\eqref{gfunction}. The normalization with $(\langle\hat{n}_1\rangle^m\langle\hat{n}_2\rangle^m)^{1/2}$ guarantees that the coherence function~\eqref{gfunction} is equal to one for a product of coherent states at each site. In general, the correlation function contains powers of the bosonic non-diagonal current $\hat{s}^{+}\vert_{\theta=0}=\hat{b}_1^\dag\hat{b}_2$ which means it measures tunneling events of $m$ particles simultaneously tunneling from site $2$ to site $1$. The index $m< N+1$ covers all non-diagonal correlation functions of the $(N+1)$-dimensional Hamiltonian in Eq.~\eqref{Lipkinmeshkovglick}, and so functions of the form~\eqref{gfunction} contain static information of the ground-state of Eq.~\eqref{Lipkinmeshkovglick} that are relevant for transport \cite{Zache2020,Schwinger1951,Schwinger1951a}. In the following sections we use analytical and numerical techniques to analyze the coherence functions in great detail. In addition, we also study spin squeezing in the anyonic Hubbard dimer in Appendix~\ref{App:squeezing}. 

\subsection{ Entangled Coherent States\label{sec:meanfield}}

For small $U$ and very large $N$ the ground-state of the Bose-Hubbard dimer is well described by a two-site Heisenberg-Weyl coherent state \cite{Mazzarella2011}
\begin{align}
	\vert\gamma(0)\ket\approx\vert\mathrm{HW}\ket=e^{\sqrt{N/2}\left(\hat{b}^{\dagger}_{1}+\hat{b}^{\dagger}_{2}-\mathrm{H.c.}\right)}\vert 0,0\ket\label{meanfieldstate1},
\end{align}
where $\vert n_1,n_2\ket$ is the state with $n_1$ particles at site 1 and $n_2$ particles at site $2$.  
The state $\vert\mathrm{HW}\ket$ describes both sites as independent coherent states with mean occupation $N/2$. This approximation implies $g^{(m)}(0)=1$ for all $m$. For $\theta\neq0$, we need to find the ground-state of the anyonic dimer which can be obtained using Eq.~\eqref{Howtogetthegroundstate} to arrive at
\begin{align}
	\vert\gamma(\theta)\ket\approx\hat{U}(\theta)\vert\mathrm{HW}\ket.\label{meanfieldstate2}
\end{align}   
The state in Eq.~(\ref{meanfieldstate2}) is an entangled coherent state~\cite{Sanders1999,Rice2000}, where the duality $\hat{U}$ entangles the product-state in Eq.~\eqref{meanfieldstate1}.
With this result we can now calculate
\begin{align}
	g^{(m)}(\theta)&\approx\bra\mathrm{HW}\vert e^{-i\theta\left[m\hat{n}_1-\binom{m}{2}\right]}\vert\mathrm{HW}\ket\nonumber\\
	&=e^{-\frac{N}{2}\left[1-e^{-im\theta}\right]+i\theta\binom{m}{2}}
\end{align}
where we used Eq.~\eqref{anyonicladdersu2} for $-\theta$, and the coherent-state properties of Eq.~\eqref{meanfieldstate1}. Details of this derivation are shown in App.~\ref{App:Meanfield}.

Here $g^{(m)}$ corresponds to a Poissonian characteristic function that embodies quantum fluctuations beyond Eq.~\eqref{meanfieldstate1}, with shifted random variable $m\hat{n}-\binom{m}{2}$, mean $N/2$ and parameter $\theta$. So the absolute value \begin{align} | g^{(m)}|=e^{-\frac{N}{2}[1-\cos(m\theta)]},
	\label{Heisenbergweylgfunction}
\end{align} is equal to one for $m\theta=2\pi n$, $n\in\mathbb{Z}$, and exponentially suppressed for all other values of $\theta$.
In fact, the behavior of $g^{(m)}$ is intimately related to the factorizability of the underlying state and so to quantum fluctuations or non-classicality. For a factorized state we would find $g^{(m)}=1$ for all $m$. The result in Eq.~\eqref{Heisenbergweylgfunction} shows that the state is only factorizable for $\theta=0$. Thus, although this theory is based on a factorized  Heisenberg-Weyl coherent state, the mean-field anyonic state~\eqref{meanfieldstate2} includes non-trivial correlations between both sites. This highlights the 
statistical interactions of anyons that result in selective filters or amplifiers of multi-particle tunneling processes. The sinusoidal dependence in Eq.~\eqref{Heisenbergweylgfunction} on the statistical parameter is thereby functionally reminiscent to the point of quasi-momentum divergence of one-dimensional anyons, found for larger system sizes in the high-density, mean-field regime~\cite{Keilmann2011,Greschner,Tang}. 

\subsection{Free Anyons \label{sec:squeezing}}
We will now move away from the mean-field, large $N$ approximation and study the coherence function $g^{(m)}$ for arbitrary finite values of $N$. In this situation, higher orders in $m$ are usually suppressed in comparison to lower orders in $m$ since the actual ground-state is not a coherent state~\cite{DellAnna:2022}. In this section we derive analytical results for $U=0$, the case of ``free'' anyons.

For $U=0$ we can adopt the spin language and realize that the ground-state of Eq.~\eqref{Lipkinmeshkovglick} is the polarized spin state pointing along the ${\hat{s}^x\vert_\theta=(\hat{s}^+\vert_\theta+\hat{s}^-\vert_\theta)/2}$ axis.
Since the state $\vert 0,N\ket$ points along the $-\hat{s}^z\vert_\theta$ direction we can then find the $\hat{s}^x\vert_\theta$-polarized state by a $\pi/2$ rotation around the ${\hat{s}^y\vert_\theta=-i(\hat{s}^+\vert_\theta-\hat{s}^-\vert_\theta)/2}$ axis. Using that the state $\vert 0,N\ket$ is invariant under multiplication of $\hat{U}$ we can then find the ground-state
\begin{align}
	\vert\gamma(\theta)\ket= e^{i\frac{\pi}{2}\hat{s}^{y}\vert_\theta} \vert 0,N\ket=\hat{U}(\theta)e^{i\frac{\pi}{2}\hat{s}^{y}\vert_{\theta=0}} \vert 0,N\ket. \label{coherentstatesanyon}
\end{align}  
Note that Eq.~\eqref{coherentstatesanyon} can be used to generate entangled $su(2)$ states, but has not been associated with a particular particle model so far \cite{Wang2000}. With this expression one can calculate analytically the $m$-th order coherence in Eq.~\eqref{gfunction} which is given by
\begin{align}\label{gfunctionsu2}
	g^{(m)}(\theta)=&\frac{N!e^{-i\frac{\theta}{2}m(N-1)}}{(N-m)!N^m}\cos^{N-m}\left(\frac{m\theta }{2}\right).
\end{align}
The details of this derivation are given in App.~\ref{App:freeanyons}.
As a first observation we find that the absolute value 
$\vert g^{(m)}(\theta)\vert$
exhibits maxima for values $m\theta=2\pi n$, $n\in\mathbb{Z}$. This is in perfect agreement with the result shown in Eq.~\eqref{Heisenbergweylgfunction}. However, in contrast to Eq.~\eqref{Heisenbergweylgfunction}, we find that $|g^{(m)}(\theta_0)|=N!/[(N-m)!N^m]\leq 1$ at $\theta_0=2\pi n/m $. This is a finite $N$ effect and for $N\to\infty$ we recover the mean-field result $|g^{(m)}(\theta_0)|\to1$. More significantly, we find that $g^{(m)}(\theta_1)=0$ for $\theta_1=\pi/m+2\pi n/m$, a result which cannot be extracted from mean-field theory in Eq.~\eqref{Heisenbergweylgfunction} which only describes the large $N$ limit. In fact, this finding means that $m$-particle tunneling events are perfectly suppressed independent of the particle number due to deconstructive interference of the tunneling phases. In the following we will study if these effects are robust against finite interactions, i.e. $U>0$.

\subsection{Role of finite $U$ \label{sec:squeezing2}}
To understand the effects of finite and large repulsive interactions $U$, we distinguish between three regimes that are adopted from the Bose dimer~\cite{Gati2007}: (i) the \textit{Rabi} regime, $U/J\ll N^{-1}$, (ii) the \textit{Josephson} regime , $N^{-1}\ll U/J\ll N$, and (iii) the \textit{Fock} regime, $N\ll U/J$. 

(i) In the \textit{Rabi} regime the bosons in the Bose dimer are basically non interacting. As a consequence we recover to a good approximation the same results as for free bosons. To move to the anyons we have to apply the unitary in Eq.~\eqref{duality} which implies that we find the ``free anyon'' situation that was discussed in the previous Sec.~\ref{sec:squeezing}.

\begin{figure}[t]
 	\center
 	\includegraphics[width=1\linewidth]{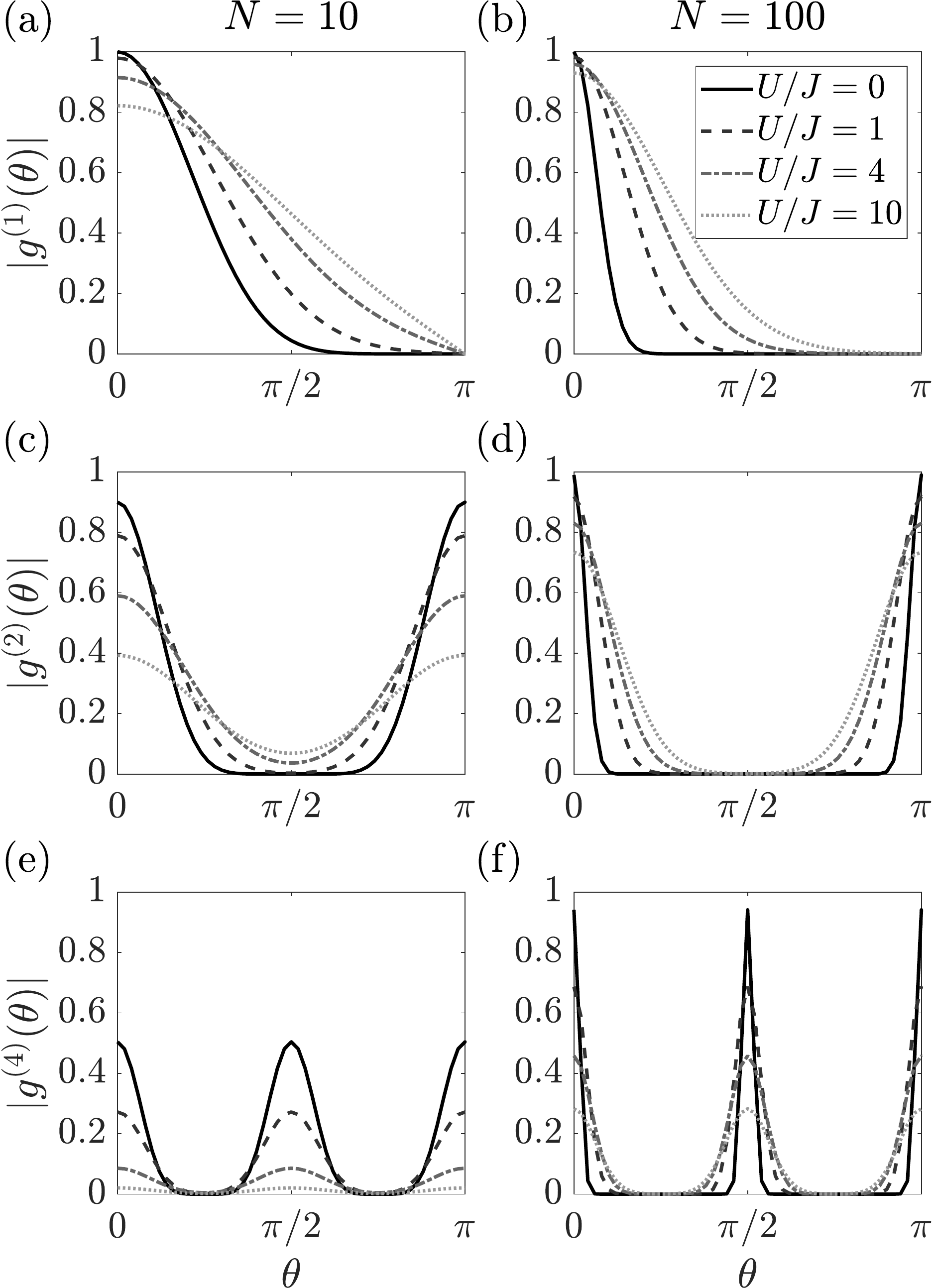}
 	\caption{The $g^{(m)}$ function calculated using Eq.~\eqref{gfunction} after finding the ground-state of Eq.~\eqref{Lipkinmeshkovglick} numerically as function of $\theta$. We show $g^{(m)}$  for $m=1$ (a), (b), $m=2$ (c), (d), and $m=4$ (e), (f) using $N=10$ (a), (c), (e) and $N=100$ (b), (d), (f). The values of $U$ in units of $J$ are visible in different gray scales [see inset of (b)] reaching from $U/J=0$ (black, solid), $U/J=1$ (dark gray, dashed), $U/J=4$ (gray, dashed-dotted), $U/J=10$ (light gray, dotted).\label{Fig:1}}
 \end{figure}

(ii) To study the effect of finite and relevant interactions $U/J>N^{-1}$ we find the ground-state $\vert\gamma(\theta)\ket$ of the Hamiltonian in Eq.~\eqref{Lipkinmeshkovglick} numerically and use it to calculate the coherence function~\eqref{gfunction}. In Fig.~\ref{Fig:1} we show the $m$-th order coherence for parameters that correspond to the \textit{Josephson} regime. In particular we show the coherence function for $m=1$ [Fig.~\ref{Fig:1}(a),(b)], $m=2$ [Fig.~\ref{Fig:1}(c),(d)],
$m=4$ [Fig.~\ref{Fig:1}(e),(f)], for $N=10$ [Fig.~\ref{Fig:1}(a),(c),(e)], and $N=100$ [Fig.~\ref{Fig:1}(b),(d),(f)]. We focus here on rather small values of $m$ since we believe that an actual measurement of the corresponding coherence functions $g^{(m)}$ becomes more challenging for larger values of $m$. Results obtained for different values of $U/J$ are shown in different gray scales [see inset of Fig.~\ref{Fig:1}(b)] including $U/J=0$ (black, solid), $U/J=1$ (dark gray, dashed), $U/J=4$ (gray, dashed-dotted), $U/J=10$ (light gray, dotted). 
The $g^{(m)}$ function for $U/J=0$ is shown as a reference and agrees with the analytical result given by Eq.~\eqref{gfunctionsu2}. Compared to this result we obtain that the coherence $|g^{(m)}|$ for $U\neq0$ is more depleted close to the original maxima $\theta_0=2\pi n/m$, $n\in\mathbb{Z}$. Close to the minima at $\theta_1=\pi/m+2\pi n/m$, instead, we find that a larger value of $U$ leads to an increase of $|g^{(m)}|$. Both features, the increase close to the minima and the decrease close to the maxima, are less pronounced for larger values of $N$. Remarkably, even for very large interaction strength $U$, we find the same oscillatory characteristics of the $|g^{(m)}|$ functions as in the $U=0$ limit. In particular, the $\theta_0$ and $\theta_1$ values of the maxima and minima are not modified. This highlights the robustness of these features whose origin lies in the non-linear tunneling phase acquired by the anyon particles. Moreover, it also shows that effects arising from the statistical parameter $\theta$ can be clearly distinguished from effects of repulsive onsite interactions. 

\begin{figure}[t]
	\center
	\includegraphics[width=1\linewidth]{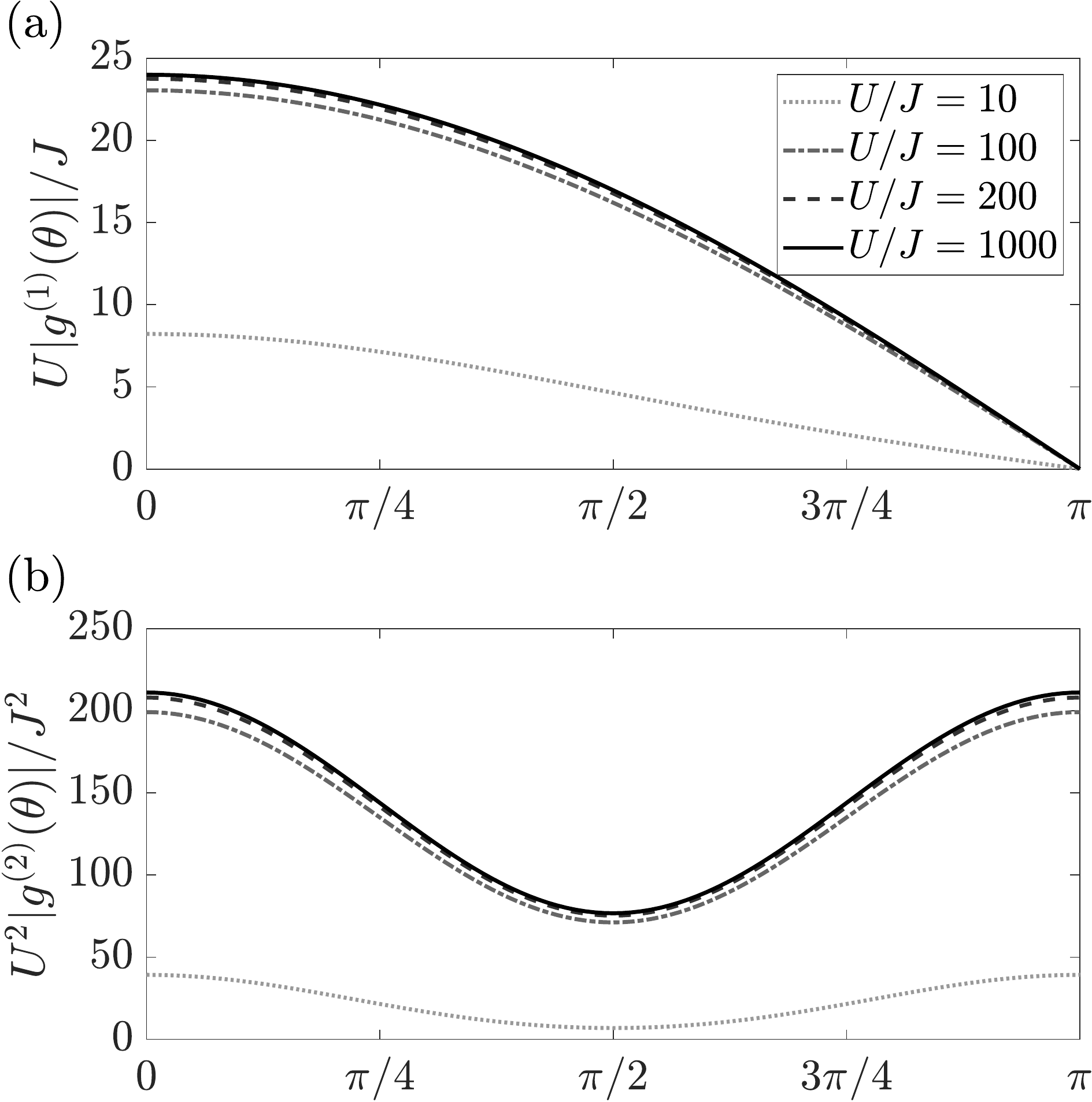}
	\caption{The $g^{(m)}$ function scaled by $(J/U)^m$ calculated using Eq.~\eqref{gfunction} after finding the ground-state of Eq.~\eqref{Lipkinmeshkovglick} numerically as function of $\theta$. We show $g^{(m)}$  for $m=1$ (a) and $m=2$ (b) for $N=10$. The values of $U$ in units of $J$ are visible in different gray scales [see inset of (a)] including $U/J=1000$ (black, solid), $U/J=200$ (dark gray, dashed), $U/J=100$ (gray, dashed-dotted), $U/J=10$ (light gray, dotted).\label{Fig:2}}
\end{figure}

(iii) We now move to the case of very strong interactions $U/J>N$, the \textit{Fock} regime. In this case the onsite repulsion prevents the emergence of coherence in the system as the ground-state of the Bose dimer is close to the state $\vert\gamma(0)\ket\approx|N/2,N/2\ket$. In fact, $m$-particle tunneling events are suppressed by a perturbative factor scaling with $(J/U)^m$. This behavior is also visible in the anyons as we check numerically: in Fig.~\ref{Fig:2} we show (a) $g^{(1)}$ and (b) $g^{(2)}$ for very large interaction strengths $U/J=10$ (light gray, dotted), $U/J=100$ (gray, dashed-dotted), $U/J=200$ (dark gray, dashed), and $U/J=1000$ (black, solid) and $N=10$.   
The $g^{(m)}$ functions for $m=1,2$ are scaled by $(J/U)^m$ to observe their convergence for large values $U/J$. This convergence is clearly visible as the curves approach a finite limit for $U/J\to\infty$. Although we find the general suppression of coherence due to $U$, the oscillatory behavior of the $g^{(m)}$ functions is still visible. This finding is also in agreement with the perturbative result
\begin{align}
g^{(1)}(\theta)\approx\frac{2J(N+2)}{U}\cos\left(\frac{\theta}{2}\right)e^{-i\frac{\theta}{2}(N-1)}	\label{eq:g1largeU}
\end{align}
that we derive in App.~\ref{App:g1largeU}. This result highlights the non-trivial dependence on $\theta$ that survives even for quite large interaction strengths.
 \section{Conclusions\label{sec:Conc}}

Statistical interactions are fundamentally different from ordinary multi-particle potentials \cite{Kundu1999,Girardeau2006,Bonkhoff2021}. For the anyon-Hubbard dimer these differences can be explained via duality to different forms of squeezing mechanisms well established in the field of quantum optics \cite{Kitagawa1993,Kitagawa1986,Sanders1999,Rice2000}. This opens a pathway to realize the physics of the anyon-Hubbard dimer in quantum optics experiments that use non-linear effects emerging from light-matter interactions~\cite{Longhi2011,Longhi2012,Yuan2017}. Remarkably, phenomena such as squeezing can be re-investigated from the perspective that the underlying particles are anyons. We find unconventional properties of higher order, bosonic correlation functions, 
similar to the non-trivial behavior of spin correlation functions in fermionic bases 
\cite{Lieb1961}. 
This is in stark contrast to the uniform destruction of coherence due to conventional interactions found for larger system sizes \cite{Keilmann2011,Greschner,Straeter2016,Tang,Bonkhoff2021}. The statistical interactions act as selective filters or amplifiers of multi-particle tunneling processes whose origin is constructive and deconstructive interference of tunneling amplitudes, respectively.  Moreover, the resulting maxima and minima are remarkably robust against onsite repulsive interactions described by $U$. This highlights the distinct character of the statistical interactions originating from the anyonic nature and onsite repulsions. 

In future it would be interesting to understand if one can apply similar non-linear transformation onto the ground state of the Bose-Hubbard model to determine or approximate the ground state of the anyonic Hubbard model with several sites. This is challenging as the unitary transformation that maps the anyonic Hubbard dimer onto the Bose-Hubbard dimer used in this paper cannot be generalized to several sites in a straightforward manner.
\begin{acknowledgments}
This work was supported by the Deutsche Forschungsgemeinschaft (DFG, German Research Foundation) via the Collaborative Research Center SFB/TR185 (Project No. 277625399) and the Research Unit FOR 2316 (Project No. P10). The authors thank Angela Foerster, Nathan Harshman, Jon Links and Luca Salasnich for valuable discussions. 
\end{acknowledgments}

\appendix

\section{Bethe Solution \label{bethesolution}}
The Bose-Hubbard dimer is an integrable sub-case of the $L$-site system, that is solvable by means of the algebraic Bethe ansatz \cite{Enolskii1993,Jon2006}. With the help of the duality relation $\hat{U}$ in  Eq.~(\ref{duality}) we have also found an exact solution for the anyonic version in Eq.~(\ref{Hamiltonian}). The dimensionless energy of the repulsive Bose-Hubbard dimer reads \cite{Jon2006}, 
\begin{align}
	E(\eta)=&-\left[\eta^{-2}\prod_{j=1}^{N}\left(1+\frac{\eta}{v_i}\right)-\frac{\eta^2 N^2}{4}-\eta^{-2}\right]\label{betheenergy},
\end{align}
with the scale $J\equiv1$ and a vanishing spectral parameter \cite{Jon2006}.
Equation~\eqref{betheenergy} only depends on the parameter $\eta=\sqrt{U/2}$ and the Bethe roots $v_j$, which have to be calculated by solving the set of algebraic equations, 
\begin{align}
	\eta^2(v^2_k)=\prod_{j\neq k}^{N}\frac{v_k-v_j-\eta}{v_k-v_j+\eta},\quad\quad k=1,...,N.\label{betheroots}
\end{align}
This yields finally the un-normalized Bethe eigenstates for the anyonic dimer by the inverse duality transform of the fundamental modes in Eq.~(\ref{inversedual1}), i.e. 
\begin{align}
	\vert\Psi\ket_{\mathrm{AH}}\propto\prod_{j=1}^{N}&\Big[\hat{b}^{\dagger}_1e^{i\theta\hat{n}_1}\left(v_j+\eta    \hat{n}_2\right)+\eta^{-1}\hat{b}^{\dagger}_2\Big]\vert0,0\ket.\label{bethesstate}
\end{align}
Having classified the models integrability, we connect the Bethe ground-state in Eq.~\eqref{bethesstate} for weak on-site interactions $U$ to the $U=0$ limit of $\hat{H}$, introduced in Sec.~\ref{sec:squeezing}. This is achieved by the asymptotic form of Bethe roots discussed in Ref.~\cite{Zhou2003a}.   
Namely, it holds that $\eta^2v_j^2\rightarrow1$ for vanishing $U$, and thus all ground-state roots become approximately equal to $v_j\approx\pm\eta^{-1}$. This ultimately violates the effective Pauli-principle for Bethe integrable bosons found by Korepin and Izergin \cite{Izergin1982}. Consequently the Bethe state converges to the non-interacting case visible in Eq.~\eqref{coherentstatesanyon}.

\section{Mean-field limit of $g^{(m)}$ \label{App:Meanfield}}

In this Appendix we show details for the derivation of Eq.~\eqref{Heisenbergweylgfunction}. First we use Eq.~\eqref{meanfieldstate2} to derive
\begin{align}
\bra\gamma(\theta)\vert \left(\hat{s}^{+}\vert_{\theta=0}\right)^m\vert\gamma(\theta)\ket=&\bra \mathrm{HW}\vert\left(\hat{s}^{+}\vert_{-\theta}\right)^m\vert \mathrm{HW}\ket
\end{align}
with the help of $\hat{U}^\dag(\theta)\hat{s}^{+}\vert_{\theta=0}\hat{U}(\theta)=\hat{s}^{+}\vert_{-\theta}$. We use then the bosonic commutation relations to calculate
\begin{align}
	\left(\hat{b}_1^\dag\hat{b}_2e^{-i\theta\hat{n}_1}\right)^m=\left(\hat{b}_1^\dag\right)^me^{-im\hat{n}_1\theta_1+i\theta\binom{m}{2}}\hat{b}_2^m.
\end{align}
Since the $\vert HW\ket$ is a product of two coherent states we can now directly apply the bosonic creation and annihilation operators on it to find 
\begin{align}
	\label{eq:gmalmost}
	g^{(m)}(\theta)&\approx\bra\mathrm{HW}\vert e^{-i\theta\left[m\hat{n}_1-\binom{m}{2}\right]}\vert\mathrm{HW}\ket.
\end{align}
Here we have used the normalization with $S=N/2$ visible in Eq.~\eqref{gfunction}. Since the operator in Eq.~\eqref{eq:gmalmost} only depends on $\hat{n}_1$ it is now sufficient to calculate
\begin{align}
\bra \sqrt{S}\vert e^{-i\theta m\hat{n}_1}\vert \sqrt{S}\ket=&e^{-S}\sum_{n=0}^\infty\frac{S^{n}}{n!}e^{-i\theta mn}\\
=&\exp\left[-S\left(1-e^{-i\theta m}\right)\right],
\end{align}
where we used the coherent state $\vert \sqrt{S}\ket$ in Fock representation $\vert \sqrt{S}\ket=\sum_{n=0}^\infty e^{-S/2}\sqrt{S^{n}/n!}|n\ket$. Applying this result in Eq.~\eqref{eq:gmalmost} leads to the formula given by Eq.~\eqref{Heisenbergweylgfunction}.

\section{Calculation of $g^{(m)}$ for free anyons \label{App:freeanyons}}
In this Appendix we present the derivation of Eq.~\eqref{gfunctionsu2}. To derive this formula we first present a representation of the spin state polarized along the $\hat{s}^x\vert_{\theta=0}$ direction. It is given by
\begin{align}
e^{i\frac{\pi}{2}\hat{s}^{y}\vert_{\theta=0}} \vert 0,N\ket=\sum_{n=0}^N\sqrt{\frac{\binom{N}{n}}{2^N}}|n,N-n\ket.
\end{align} 
Therefore we find the ground-state of the anyon dimer for $U=0$ to be
\begin{align}
	\vert\gamma(\theta)\ket=\sum_{n=0}^Ne^{i\frac{\theta}{2}n(n-1)}\sqrt{\frac{\binom{N}{n}}{2^N}}|n,N-n\ket.
\end{align}
We can now use $\hat{s}^+|_{\theta=0}=\hat{b}_1^\dag\hat{b}_2$ to calculate
\begin{align}
	&[\hat{s}^+|_{\theta=0}]^m|n,N-n\ket\nonumber\\&=\sqrt{\frac{(n+m)!(N-n)!}{n!(N-n-m)!}}|n+m,N-n-m\ket.
\end{align}
With this result we obtain
\begin{align}
	\bra[\hat{s}^+|_{\theta=0}]^m\ket
	=&\sum_{n=0}^{N-m}\frac{N!e^{i\frac{\theta}{2}\left[-m(m-1)-2(n-m)m\right]}}{2^N n!(N-n-m)!}.
\end{align}
Using the series expansion
\begin{align}
	\sum_{n=0}^{N-m}\frac{N!}{n!(N-n-m)!}X^{n-m}
	=&\frac{N!}{(N-m)!}(1+X)^{N-m}
\end{align}
we get
\begin{align}
\bra[\hat{s}^+|_{\theta=0}]^m\ket=&\frac{N!e^{-i\frac{\theta}{2}m(m-1)}}{2^N(N-m)!}\left(1+e^{-i\theta m}\right)^{N-m}\nonumber\\
=&\frac{N!e^{-i\frac{\theta}{2}m(N-1)}}{2^m(N-m)!}\cos^{N-m}\left(\frac{m\theta}{2}\right).
\end{align}
To calculate $g^{(m)}$ we use Eq.~\eqref{gfunction} where we need to divide by $S^m=N^m/2^m$. This leads to the result which is presented in Eq.~\eqref{gfunctionsu2}.
\section{Calculation of $g^{(1)}$ for large $U$ \label{App:g1largeU}}
In this section we calculate the $g^{(1)}$ function for arbitrary $\theta$ but large $U\gg JN$.
For this we first calculate the perturbative ground-state of the Bose dimer and use then the unitary transformation in Eq.~\eqref{duality} to calculate the anyon ground-state. 

For the Bose dimer in the limit ${U\to\infty}$ the ground-state is given by
\begin{align}
	\vert\gamma_0(0)\ket=\vert N/2,N/2\ket.
\end{align}
For large but finite $U$ we can calculate the ground-state in first-order perturbation theory such that $\vert\gamma(0)\rangle\approx\vert\gamma_0(0)\rangle+\vert\gamma_1(0)\rangle$ with
\begin{align}
\vert\gamma_1(0)\rangle=\frac{2J\sqrt{\left(\frac{N}{2}+1\right)\frac{N}{2}}}{U}\left(\vert 1\ket+\vert-1\ket\right),
\end{align}
where we used the short notation $\vert l\ket=\vert N/2+l,N/2-l\ket$.
In a next step we can apply $\hat{U}(\theta)$ [Eq.~\eqref{duality}] to find
$\vert\gamma(\theta)\rangle\approx\vert\gamma_0(\theta)\rangle+\vert\gamma_1(\theta)\rangle$ with $\vert\gamma_0(\theta)\rangle=\hat{U}(\theta)\vert\gamma_0(0)\rangle$ and $\vert\gamma_1(\theta)\rangle=\hat{U}(\theta)\vert\gamma_1(0)\rangle$. The calculation yields 
\begin{align}
	\vert\gamma_0(\theta)\ket=&e^{i\frac{\theta}{2}\frac{N}{2}\left(\frac{N}{2}-1\right)}\vert 0\ket,\\
	\vert\gamma_1(\theta)\rangle=&\frac{2J\sqrt{\left(\frac{N}{2}+1\right)\frac{N}{2}}}{U}e^{i\frac{\theta}{2}\frac{N}{2}\left(\frac{N}{2}+1\right)}\vert 1\ket\nonumber\\
	&+\frac{2J\sqrt{\left(\frac{N}{2}+1\right)\frac{N}{2}}}{U}e^{i\frac{\theta}{2}\left(\frac{N}{2}-1\right)\left(\frac{N}{2}-2\right)}\vert-1\ket.
\end{align}
Using this result we can now evaluate
\begin{align}
	g^{(1)}(\theta)\approx\frac{\bra \gamma_1(\theta)\vert\hat{b}_1^\dag\hat{b}_2\vert\gamma_0(\theta)\ket+\bra \gamma_0(\theta)\vert\hat{b}_1^\dag\hat{b}_2\vert\gamma_1(\theta)\ket}{N/2}
\end{align}
which results in Eq.~\eqref{eq:g1largeU}.
\section{Spin squeezing in the Anyonic Hubbard dimer \label{App:squeezing}}
In this section we briefly discuss spin squeezing in the anyonic Hubbard dimer. We use the definition of spin squeezing in a broader sense that measures entanglement within an ensemble of spin-$1/2$ particles. Namely, in our setting each boson can occupy two sites which makes it effectively a spin-$1/2$ particle. To show the existence of spin squeezing for many particles we  calculate the variance of the collective spins 
\begin{align}
(\Delta \hat{s}^{a}\vert_{\theta=0})^2=\langle\gamma(\theta)|(\hat{s}^{a}\vert_{\theta=0})^2|\gamma(\theta)\rangle-\langle\gamma(\theta)|\hat{s}^a\vert_{\theta=0}|\gamma(\theta)\rangle^2\label{eq:var}
\end{align}
for the spins $\hat{s}^x\vert_{\theta=0}=(\hat{b}_1^\dag\hat{b}_2+\hat{b}_2^\dag\hat{b}_1)/2$, $\hat{s}^y\vert_{\theta=0}=i(\hat{b}_2^\dag\hat{b}_1-\hat{b}_1^\dag\hat{b}_2)/2$, $\hat{s}^z\vert_{\theta=0}=(\hat{n}_1-\hat{n}_2)/2$, $a=x,y,z$, and with the help of Eq.~\eqref{Howtogetthegroundstate}.

For pure states the variances are directly connected to the quantum Fisher information which is given by
\begin{align}
\mathcal{F}_a=4(\Delta \hat{s}^{a}\vert_{\theta=0})^2.\label{eq:Fisher}
\end{align}
The quantum Fisher information $\mathcal{F}_a$ is a witness for metrological useful entanglement and determines the usefulness of a quantum state in estimating a small phase shift $\varphi$ that generates the rotation $\exp(-i\varphi \hat{s}^a\vert_{\theta=0})$~\cite{Pezze:2018}. In that sense $\mathcal{F}_a>N$ implies useful quantum entanglement within the ensemble of spin-$1/2$ particles and $\mathcal{F}_a=N^2$ is the fundamental upper bound called Heisenberg limit. This connection to quantum metrology makes the quantum Fisher information a prominent tool for quantifying entanglement in spin systems where $\mathcal{F}_a>N$ implies spin squeezing. 

\begin{figure}[h!]
	\center
	\includegraphics[width=1\linewidth]{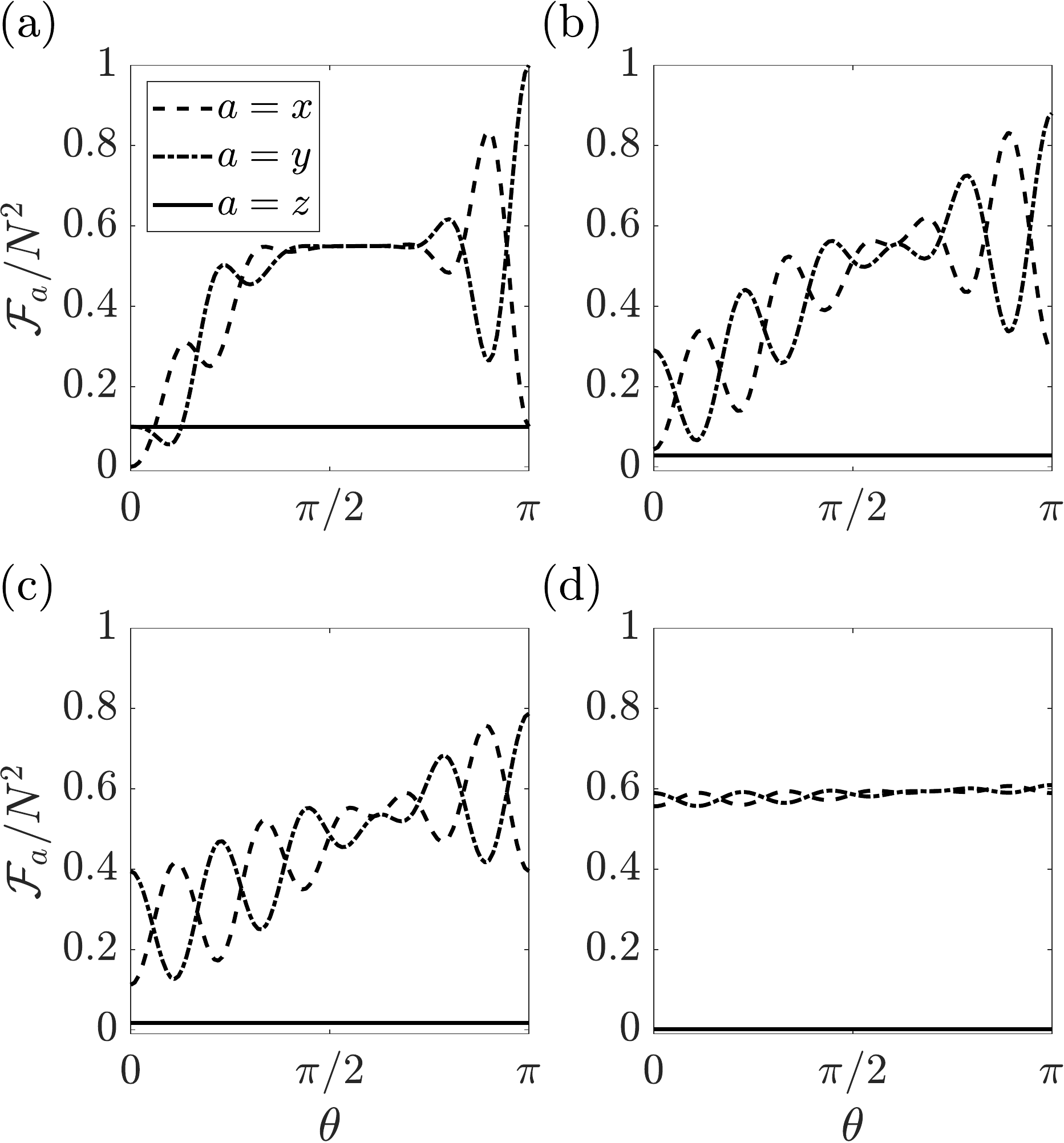}
	\caption{Quantum Fisher information $\mathcal{F}_a$ normalized by $N^2$ calculated using Eqs.~\eqref{eq:var} and \eqref{eq:Fisher} as function of $\theta$ for (a) $U=0$, (b) $U=J$, (c) $U=10J$, and (d) $U=100J$. Different line styles correspond to the different variances $a=x,y,z$ [see inset of (a)]. The particle number is $N=10$.\label{Fig:S1}}
\end{figure}

In Fig.~\ref{Fig:S1} we plot the quantum Fisher information for various values of $\theta$ and onsite interactions strengths (a) $U=0$, (b) $U=J$, (c) $U=10J$, and (d) $U=100J$. In all subfigures we observe that $\mathcal{F}_z$ is constant as a function of $\theta$. This is a consequence of $\hat{s}^z$ being invariant under the transformation $\hat{U}(\theta)$. The value of $\mathcal{F}_z$ is reduced for increasing values of $U$ which shows that larger values of $U$ suppress onsite particle number fluctuations. In Fig.~\ref{Fig:S1}(a) we observe $\mathcal{F}_x\approx N$ and $\mathcal{F}_y\approx N$  for small values of $\theta\approx 0$. For larger values of $\theta$ we find $\mathcal{F}_x>N$ and $\mathcal{F}_y>N$ for a broad range of $\theta$ values which highlights the existence of spin squeezing originating from the transformation $\hat{U}(\theta)$. At $\theta=\pi$ the quantum Fisher information approaches the Heisenberg limit, i.e. $\mathcal{F}_y=N^2$. This is in full agreement with one-axis twisting~\cite{Kitagawa1993,Pezze:2018}.

 In Fig.~\ref{Fig:S1}(b), (c) we find that at $\theta=0$ a non-vanishing value of $U$ already results in spin squeezing $\mathcal{F}_x>N$ and $\mathcal{F}_y>N$. Remarkably we clearly see changes in $\mathcal{F}_x$ and $\mathcal{F}_y$ when varying $\theta$ which allows to discriminate entanglement effects originating from density-dependent tunneling and onsite repulsion. Only for very large values of $U=100$ we see a rather constant behavior of $\mathcal{F}_x$ and $\mathcal{F}_y$ which is visible in Fig.~\ref{Fig:S1}(d).

\bibliographystyle{apsrev4-2}       
\bibliography{RefsAnyon.bib}

\end{document}